\documentclass[aps,prbbib,twocolumn,epsf]{revtex4}
\usepackage{graphicx}

\def\beq{\begin{eqnarray}}
\def\eeq{\end{eqnarray}}

\begin{document}
\title{The maximum density droplet to lower density droplet transition in quantum dots}

\author{A. D. G\"u\c{c}l\"u}
\affiliation{Theory Center, Cornell University, Ithaca, NY 14853}
\author{C. J. Umrigar}
\affiliation{Theory Center and Laboratory of Atomic and Solid State Physics, Cornell University,
 Ithaca, NY 14853}

\date{\today}
\begin{abstract}
We show that, Landau level mixing in two-dimensional quantum dot wave functions can be taken
into account very effectively by multiplying
the exact lowest Landau level wave functions by a Jastrow factor which is optimized by
variance minimization. The comparison between exact diagonalization
and fixed phase diffusion Monte Carlo results suggests that the phase
of the many-body wave functions are not affected much by Landau level mixing.
We apply these wave functions to study the transition from the maximum
density droplet state (incipient integer quantum Hall state with angular momentum $L=N(N-1)/2$)
to lower density droplet states $(L>N(N-1)/2$).
\end{abstract}


\maketitle

\section{Introduction}

Advances in miniaturization and processing
techniques of semiconductor systems have made it possible to create
nanostructures containing only a small number of
electrons~\cite{jacak_1998, ashoori_1996,ashoori_1992,ashoori_1993,kouwenhoven_1997,oosterkamp_1999}.
These quantum dots, also called artificial atoms, provide a
crucial testing ground for many important quantum effects in low-dimensional systems.
%that were largely unexplored.
The system parameters can be tuned
by controlling the dot geometry, electrostatic gate voltage,
and by applying a magnetic field.

Experimental techniques such as capacitance spectroscopy~\cite{ashoori_1992,ashoori_1993}
or gated transport spectroscopy~\cite{kouwenhoven_1997,oosterkamp_1999}
allow experimentalists to precisely control the number of
electrons in a two-dimensional parabolic confining potential, and map their electronic
properties as a function of the magnetic field. Such systems offer very rich quantum many-body effects,
providing a link between two different branches of physics: In the limit of zero magnetic field
the system acts like an atom, i.e., it exhibits shell structure and obeys Hund's first rule.  On the
other hand, for relatively modest magnetic fields the system exhibits incipient quantum
Hall physics.  A large number of transitions
induced by the magnetic field can be observed experimentally.  The transitions
can be understood from Hartree-Fock theory or from density functional theory
(at low magnetic fields even the simple constant interaction model~\cite{McEuen92} suffices)
but at high magnetic fields an accurate treatment of many-body effects is important.

The aim of this work is to study in detail the transitions between many-body states for moderately
large magnetic field values where the angular momentum of the system increases beyond that of the
so called maximum density droplet (MDD) regime~\cite{oosterkamp_1999},
corresponding to the filling factor $\nu=1$ in quantum Hall physics. Most of the
previous work on the transition between MDD to lower density states (LDD) was performed in
the lowest Landau level (LLL) approximation and/or was restricted to spin polarized electrons
~\cite{eric_yang_1993,hawrylak_1993,martin-moreno_1995,wojs_1997,goldman_1999,manninen_2001}.
The LDD states are not always predicted to be fully spin polarized, both in the
absence of Landau level (LL) mixing~\cite{eric_yang_1993} and in the presence
of LL mixing ~\cite{siljamaki_2002}.

Here, we combine exact diagonalization and quantum Monte Carlo (QMC)
techniques~\cite{foulkes_2001,NATObook,HammondLesterReynolds94} to obtain a very accurate description
of many-body states beyond the MDD.  Linear combinations of determinants obtained by
exact diagonalization within the LLL approximation serve as the starting point for
building trial wave functions for our
variational Monte Carlo (VMC) and diffusion Monte Carlo (DMC) calculations. The method described
here is very systematic, can be applied to any eigenstate,
and, is more efficient than exact diagonalization with LL mixing. Trial wave functions obtained
by the exact diagonalization method have the right symmetry properties, including conservation of total angular momentum $L$,
total spin $S$,
and separability of center of mass and relative motion, which allows us to study center of mass excitations
~\cite{wensauer_2004}.
We apply our method to investigate the many-body states involved in the
MDD-LDD transition. We find that for the $N=4$ electron system, the pair densities of the ground and
excited states
close to the MDD-LDD transition have a well defined square symmetry, while those for $N=6$ states have
little structure.  We find that the MDD-LDD transition acts as a convergence point for special
values of $(L,S)$ which correspond to the magic numbers~\cite{maksym_2000}.
The redistribution of electronic charge is rather smooth in the absence of the
Zeeman effect, but, when a Zeeman term is included in the Hamiltonian
it becomes more abrupt.

\section{Model and method}

\subsection{Hamiltonian}

We consider an $N-$electron two-dimensional quantum dot with a circularly
symmetric parabolic confinement potential. Within the effective
mass approximation, and neglecting finite thickness effects, the
Hamiltonian in an external magnetic
field, $B$, is
\begin{eqnarray}
H&=&\sum_j^N
 \Bigg( \frac{1}{2m^*}\left({\bf p}_j+\frac{e}{c}{\bf A}({\bf r}_j) \right)^2
 +\frac{1}{2} m^* \omega_0^2{\bf r}_j^2 \nonumber \\
&& + {e^2 \over \epsilon} \sum_{i < j}^N {1 \over r_{ij}}
 + g^*\mu_B B s_{z,j} \Bigg),
\label{H1e}
\end{eqnarray}
where $m^*$ is the effective mass of electron, ${\bf A}$ is the vector
potential, $\omega_0$ is the parameter characterizing the parabolic
confinement potential, $\epsilon$ is the dielectric constant,
$g^*$ is the effective $g$-factor, $\mu_B$ is the Bohr magneton
and $s_{z,j}$ is the $z$-component of the spin of the $j^{th}$ electron.

We use standard materials constants
corresponding to GaAs, $m^*=0.067m_0$, dielectric constant $\epsilon=12.4$
and $g^*=-0.44$.
Results will be given in effective
atomic units ($\hbar = e^2/\epsilon = c = m^* =1$):
the effective Bohr radius is $a^*_0=9.79373$ nm,
and the effective Hartree is $H^*=11.8572$ meV.
We choose the confinement parameter $\hbar\omega_0$ to be 0.28 H$^*$ or 3.32 meV,
which is within the typical experimental range
~\cite{ashoori_1993,kouwenhoven_1997,oosterkamp_1999}.

\subsection{Single-particle states}
Ignoring the interaction term, the single particle Fock-Darwin
energies of this Hamiltonian are~\cite{fock-darwin}
\beq
\epsilon_{n_r l}&=&(2n_r+\vert l \vert+1) \hbar \omega - {l \hbar \omega_c \over 2},
\label{FockDarwin}
\eeq
where $\omega_c=eB/m^* c$ is the
cyclotron frequency, $\omega=\sqrt{\omega_0^2+\omega_c^2/4}$,
$n_r$ is the radial quantum number and $l$ is the angular
quantum number ($n_r,\vert l \vert =0,1,...,\infty$).
Equivalently, the energy spectrum can be expressed as a sum of two
harmonic-oscillators~\cite{hawrylak_1993},
\beq
\epsilon_{nm}= (n+1/2)\hbar \omega_+ + (m+1/2)\hbar\omega_-,
\eeq
where $\omega_{\pm}=\omega \pm \omega_c/2$,
$n$ is the Landau level index, $n,m=0,1,...,\infty$
and $l=-n,-n+1,...,\infty$.
The relationship between these two sets of quantum numbers is
$l=m-n,\; n_r=\min(n,m)$ or
$n=n_r+{\vert l \vert -l \over 2} = n_r+\max(0,-l),\;
m=n_r+{\vert l \vert +l \over 2} = n_r+\max(0,l)$.

At zero $B$ the energies depend only on the principal
quantum number, $2n_r+\vert l \vert = n+m$, whereas at infinite $B$ they
depend only on the Landau level index $n$.
The corresponding eigenfunctions $\vert nm \sigma \rangle$
($\sigma$ is the spin index)
are used as the basis set in our exact diagonalization and QMC calculations
of interacting dots.
At zero magnetic field the system has degeneracies and therefore a
shell structure, as in the case of real atoms.
When a magnetic field is applied, the degeneracy is broken and several
transitions occur due to single-electron level crossings as described by
Eq.~\ref{FockDarwin}.  Eventually all the electrons are in lowest
Landau level orbitals and each of these orbitals is doubly occupied.
In a noninteracting electron picture no more transitions are predicted
as the magnetic field is increased further, but, experimentally they
are observed~\cite{oosterkamp_1999}.  These can be understood from
Hartree-Fock or density functional theory.  The intra Landau level splittings
get smaller with increasing $B$ and the exchange interaction
favors parallel spins, leading to spin flips, until eventually all the
spins are parallel and the $N$ lowest angular momentum states are occupied
singly.  This is the maximum density droplet (MDD) state.
As $B$ is increased further, Hartree-Fock~\cite{chamon_1994} and
density functional calculations~\cite{ferconi_1994,ferconi_1997}
predict an edge reconstruction due to some electrons jumping from low angular
momentum states to higher angular momentum states because the Coulomb interaction
overwhelms the parabolic external potential.  These lower density
droplet (LDD) states were observed experimentally\cite{klein_1995}
but quantitative discrepancies with experiments exist.
Since several determinants of LLL noninteracting orbitals are exactly
degenerate it is expected that an accurate treatment of correlation is important.
This is the range of $B$ that we study in this paper.

\subsection{Many-particle states and Exact diagonalization}
The many-body wave functions may be chosen to be eigenstates
of the total angular momentum $\hat{L} \equiv \hat{L}_z$, total spin $\hat{S}^2$ and the z-component
of total spin $\hat{S}_z$.
In addition, the dynamic symmetry of an isotropic parabolic potential leads
to the separability of center of mass and relative motion and the existence of
two {\it center of mass operators}~\cite{wensauer_2004},
$\hat{C}_+$ and $\hat{C}_-$ that
commute with the $\hat{H}$, $\hat{L}$, $\hat{S}^2$ and $\hat{S}_z$,
 and the above operators:
\begin{eqnarray}
\lefteqn{\hat{C}_+ =} \nonumber \\
\!\!\!&\!\!\!&\!\!\!\frac{1}{N}\sum_{n'm'\sigma'nm\sigma} \sqrt{(n'+1)n}
\hat{c}^\dagger_{(n'+1)m'\sigma'}\hat{c}_{n'm'\sigma'}\hat{c}^\dagger_{(n-1)m
\sigma}\hat{c}_{nm\sigma} \nonumber\\
\\
\lefteqn{\hat{C}_- =} \nonumber \\
\!\!\!&\!\!\!&\!\!\!\frac{1}{N}\sum_{n'm'\sigma'nm\sigma} \sqrt{(m'+1)m}
\hat{c}^\dagger_{n'(m'+1)\sigma'}\hat{c}_{n'm'\sigma'}\hat{c}^\dagger_{n(m-1)
\sigma}\hat{c}_{nm\sigma}. \nonumber\\
\end{eqnarray}
Determinants of Fock-Darwin orbitals are
eigenfunctions of $\hat{L}$ and $\hat{S}_z$.
%but not of $\hat{S}^2$, $\hat{C}_+$ and $\hat{C}_-$.
Thus, the many-body Hamiltonian in a determinantal basis is block diagonal,
with the blocks corresponding to different $(L,S_z)$ values.
Also, we note that LLL determinants are eigenfunctions of $\hat{C}_+$
with eigenvalues $C_+=0$.
The size of the subspaces can be further reduced by employing linear combinations
of determinants, known as configuration state functions (CSFs), that are eigenstates
of $\hat{S}^2$ and $\hat{C}_-$.
The diagonalization of $\hat{S}^2$ and $\hat{C}_-$ is done as
explained in Ref.(\onlinecite{wensauer_2004}).
The matrix elements of the Coulomb interaction can be determined either by numerical
integration or analytically~\cite{Hawrylak_1993b,WojsHawrylak_1995,Tsiper02}.

For a state with a given total angular momentum $L$, and for a given number
of Landau levels, $n$, only single-particle states with angular momentum
$l_{\rm min} \le l \le l_{\rm max}$
contribute, where $l_{\rm min}=-n$.
For fully spin polarized LLL wave functions, $l_{\rm max}=L-(N-1)(N-2)/2$
and for lower spin states it is higher than that.
As the number of Landau levels is increased, $l_{\rm max}$ increases, both because
the remaining $N-1$ electrons can lower their angular momenta by occupying more than
one Landau level and because
the $n^{th}$ Landau level has states with $l \ge -n$.
Empirically we find that determinants containing orbitals with $l > 2L/N+1$
contribute little to the wave function and consequently we could limit
ourselves to orbitals with $l \le 2L/N+1$.
However, in most cases we avoid doing this because
the wave functions are no longer exact eigenstates of the operator $\hat{C}_-$.

\subsection{Quantum Monte Carlo}
In this work, we use both the variational Monte Carlo (VMC) and the
diffusion Monte Carlo (DMC) methods~\cite{foulkes_2001,NATObook,HammondLesterReynolds94}
to improve the exact diagonalization
wave functions and energies obtained within the LLL.
First, we introduce variational parameters by
multiplying the LLL wave function by a Jastrow factor and
optimize the Jastrow parameters by minimizing the
variance of the local energy~\cite{umrigar_1988}. For a given
state $(L,S,S_z,C_-)$, the form of our trial wave functions
is
\begin{equation}
  \Psi_{T}^{L,S,S_z,C_-}=J_{ee}J_{ed}J_{eed}
    \sum_i^{N_{\rm CSF}}\alpha_i \Psi_D^{L,S,S_z,C_-},
\end{equation}
where $J_{ee}, J_{ed}, J_{eed}$ are the electron-electron, electron-dot,
and electron-electron-dot Jastrow terms respectively.
The CSF's $\Psi_D^{L,S,S_z,C_-}$ are obtained by diagonalizing the $\hat{S}^2$
and $\hat{C}_-$ matrices.
Multiplying by the Jastrow factor does not alter the $L$, $S^2$ and $S_z$ symmetries of the wave function,
aside from introducing a completely negligible amount of spin contamination~\cite{HuangFilippiUmrigar98}
(not an exact eigenstate of $\hat{S^2}$), arising from the different cusp conditions for
parallel- and antiparallel-spin electrons.
The Jastrow factor reduces the statistical error in both VMC and DMC for a given number of Monte Carlo steps.
It also lowers the VMC energy, but leave the fixed-phase DMC energy (described later) unchanged because
the phase of the trial wave function is not altered.

The linear coefficients $\alpha_i$ are obtained by diagonalizing $\hat{H}$.
They can also serve as additional free parameters that can be reoptimized in VMC.
%It is even possible to completely skip the diagonalization of
%$\hat{H}$, and optimize the linear coefficients purely within VMC.
However, we have found that, at sufficiently large magnetic fields, there is no gain in energy
from reoptimizing them, and the exact LLL wave function multiplied by the optimized Jastrow factor
provides an excellent trial wave function.  This is in contrast to the situation
in the absence of a magnetic field, where the reoptimized linear coefficients in the presence
of the Jastrow factor are considerably smaller in magnitude than the original ones from
exact diagonalization.
This can be understood as follows.
Quantum chemists distinguish between static correlation (also known as near-degeneracy correlation)
and dynamic correlation.
Within the quantum Monte Carlo literature it is well understood that static correlation
is most effectively incorporated into the wave function by using a linear combination
of nearly degenerate determinants, whereas dynamic correlation is most efficiently
treated by a flexible Jastrow factor.
In the case of dots in sufficiently large magnetic fields the separation of energy scales
(intra Landau level versus inter Landau level)
%becomes very sharp and the above ideas work exceptionally well.
is very large and the distinction between static and dynamic correlation becomes sharp.

For any given state, the determinantal coefficients $\alpha_i$, for a LLL wave function, need not be
recalculated when the system parameters $m^*$, $\omega_0$, $B$ and $\epsilon$ are varied, because
the $\alpha_i$ are independent of these system parameters.
This is because all the LLL determinants with a given $L$ have the same expectation value for the single-particle Hamiltonian.
So the single-particle terms in the Hamiltonian contribute a multiple of the unit matrix
and only the interaction part of the Hamiltonian need be diagonalized.
Hence, the LLL wave function is independent of $\epsilon$, and, depends on
$m^*$, $\omega_0$ and $B$ only through a scaling of
the linear dimensions by the factor $\sqrt{1/m^*\omega}$.
Consequently these Jastrow-LLL wave functions are not only a very accurate choice
but also a very efficient choice for QMC calculations.
Of course this does not apply to low angular momentum $L$ states because it is impossible
to construct them using only LLL orbitals.
Note also that the independence of the $\alpha_i$ on the system parameters makes the
Fock-Darwin orbitals the most efficient choice, as opposed to using say Hartree Fock
or density functional orbitals.

The DMC method improves upon the VMC energy by projecting onto
the lowest energy wave function that has the same nodes (or the same phase in the case of
complex wave functions) as the trial wave function.  The resulting fixed-node or
fixed-phase energies are upper bounds to the true energies and consequently
the accuracy of the calculations depends on the quality of the nodes or the phase
of the trial wave functions.
We employ an efficient implementation of DMC described in Ref.~\onlinecite{umrigar_1993}.
Since we use complex trial wave functions, we employ
the fixed phase approximation~\cite{ortiz_1993}, which has previously been
applied to quantum dot systems~\cite{bolton_1996,guclu_2003}:
starting from a variational wave function $\Psi_{T}$, a new function
$f=\Phi^*\Psi_{T}$ is introduced, where $\Phi$ is the ground state of the
$N$-body Hamiltonian subject to the constraint that it have the same phase as $\Psi_T$.
The integral form of the Schr\"odinger equation in imaginary
time ($t\rightarrow it$) then can be written in terms of $f$ as:
\begin{equation}
  f({\bf R},t+\tau)=\int d{\bf R'}\tilde{G}({\bf R,R'},\tau)f({\bf R'},t),
\label{mastereq}
\end{equation}
where ${\bf R}$ represents the coordinates of N electrons,
$G({\bf R,R'},\tau)= \Psi_{T}({\bf R})\langle {\bf R}\vert e^{-\tau\hat{H}}\vert {\bf R}' \rangle / \Psi_{T}({\bf R'})$,
is a Green function that is explicitly known only in the short-time limit ($\tau \to 0$).
Hence, the Eq.(~\ref{mastereq}) can be solved by applying
$G$ repeadetly until the desired projection is obtained.
Since the function $f$ is used as a probability density for Monte Carlo sampling, it must
be real and positive which is equivalent to the assumption that the phase of $\Phi$ is
the same as that of the trial wave function $\Psi_{T}$.
After equilibration, the energy can be calculated as an average over the
sampled points ${\bf R}_m$:
\begin{equation}
  E\approx\frac{1}{M}\sum_{m}^M E_L({\bf R}_m),
\end{equation}
where $E_L$ is the real part of the local energy,
\begin{equation}
  E_L({\bf R})=\mbox{Re}\left\{\Psi_{T}({\bf R})^{-1}
               \hat{H}\Psi_{T}({\bf R})\right\}.
\end{equation}

We note, however,
that for an external potential with circular symmetry it is also possible to use
the fixed node approximation: In the subspace of the good quantum number $L$,
the Hamiltonian can be written as
\begin{equation}
\hat{H}
  = \sum_j \left( \frac{1}{2m^*} {\bf p}_j^2 +
    \frac{1}{2} m^* \omega^2 {\bf r}_j^2
    + {e^2 \over \epsilon} \sum_{i\not= j}\frac{1}{r_{ij}} \right) - \frac{1}{2}\hbar\omega_c L
\end{equation}
which is purely real.
We can ignore the last term in $\hat{H}$ since it just gives a trivial shift in
the energies of the various $L$ states.  For the remaining part of $\hat{H}$, states with
angular momentum $L$ and $-L$ are degenerate and so we can consider their sum, which
is real.
It is possible to expand these real LLL wave functions in
terms of real orbitals, but the number of determinants increases greatly.
For instance, a MDD state can be written as a single determinant of complex orbitals
or a sum of $2^{N-1}$ determinants of real orbitals.
So, it is more efficient to use complex wave functions within
the fixed phase approximation.

\section{Results and discussion}

We now apply the techniques described above to quantum dots similar to those
studied experimentally~\cite{ashoori_1993,kouwenhoven_1997,oosterkamp_1999},
in magnetic fields ranging from $3-8$ Tesla bringing the dots in the vicinity
of MDD-LDD transition.
Since the Zeeman term just results in a trivial shift of the different
$S_z$ states relative to each other, we will ignore the Zeeman term unless
explicitly stated otherwise.

\subsection{Comparison between Exact diagonalization and QMC}
\begin{figure}[htb]
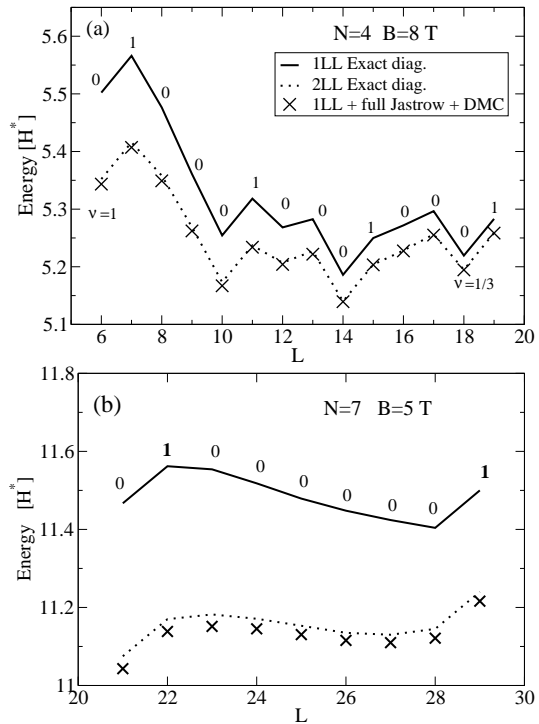

\includegraphics[width=7cm,clip]{fig1a.eps}
\includegraphics[width=7cm,clip]{fig1b.eps}
\caption{Energy as a function of total angular momentum $L$, for (a) $N=4$ at $B=8$ T,
(b) $N=7$ at $B=5$ T, obtained by
exact diagonalization in LLL approximation (solid line), exact diagonalization
including 2LLs (dotted line),
and DMC using full Jastrow term (cross marks). The numbers above the line
represent the $C_-$ quantum number.}
\label{E_L}
\end{figure}

In this subsection, we compare results obtained from exact diagonalization and QMC
calculations with different levels of approximation.  Fig.~\ref{E_L} shows the energies
of $N=4$ and $N=7$ fully spin polarized electrons at $B=8$ T and $B=5$ T respectively.
The solid line represents
the results obtained by exact diagonalization within the LLL approximation.
from which we build QMC trial wave functions.
The dotted lines are the exact
diagonalization results with 2 LLs included.
The DMC results, using the Jastrow-LLL wave functions, are represented
with the crosses. The numbers represent $C_-$, with $0$ indicating that
there is no center of mass excitation.
As seen from Fig.~\ref{E_L},
DMC energies are very accurate, better than those from a 2 LL diagonalization. This is
particularly clear in Fig.~\ref{E_L}(b) where the magnetic field is only 5 T
since LL mixing is more important at lower magnetic fields.
For 4 spin-polarized electrons
there is no real computational advantage to using QMC over exact diagonalization,
but, for spin unpolarized dots and/or higher numbers of electrons the QMC method becomes
more efficient, because the number of determinants needed is dramatically smaller.
For instance, in Fig.~\ref{E_L}(b) the QMC calculation for $L=29$
using 21 determinants in the
trial wave function clearly gives a better energy than a 2 LL diagonalization using 28637
determinants. These results also verify the appearance of magic numbers, {\it i.e.}
particular angular momenta, $L^*$, at which a deep dip occurs, followed by
a center of mass excitation at $L^*+1$~\cite{maksym_2000,jacak_1998}.
Also, we note that for $N=7$, within the LLL approximation exact diagonalization
makes an incorrect prediction for the ground state which underscores the need to go
beyond the LLL approximation near the MDD-LDD transition.

\begin{table}

\caption{Comparison of energies for different $L$ values ($N=4$, $S=2$, $B=8$ T)
obtained from exact diagonalization method within 1,2, and 3 LL approximations (ED), VMC
including only the electron-electron part of the Jastrow factor (e-e VMC), and DMC using
a fully optimized Jastrow factor (DMC). The energy differences between DMC and the
3-LL exact diagonalization, shown in the last column, are small, with the DMC energies being better in most cases.}
\begin{tabular}{ccccccc}

\hline
% L  & 1LL ED & 2LL ED & 3LL ED & e-e VMC & DMC & $E_{DMC}-E_{ED}$ \\
  L  & 1LL ED & 2LL ED & 3LL ED & e-e VMC & DMC & DMC-ED \\
\hline
  6  &   5.5022 & 5.3522 & 5.3463 & 5.3459 & 5.3434 & {\bf $-29\times 10^{-4}$} \\
  7  &   5.5660 & 5.4159 & 5.4100 & 5.4096 & 5.4071 & {\bf $-29\times 10^{-4}$} \\
  8  &   5.4764 & 5.3582 & 5.3511 & 5.3517 & 5.3491 & {\bf $-20\times 10^{-4}$} \\
  9  &   5.3609 & 5.2688 & 5.2641 & 5.2634 & 5.2622 & {\bf $-19\times 10^{-4}$} \\
 10  &   5.2544 & 5.1733 & 5.1705 & 5.1738 & 5.1702 & {\bf $-3\times 10^{-4}$} \\
 11  &   5.3182 & 5.2370 & 5.2343 & 5.2376 & 5.2339 & {\bf $-4\times 10^{-4}$} \\
 12  &   5.2683 & 5.2070 & 5.2042 & 5.2064 & 5.2041 & {\bf $-1\times 10^{-4}$} \\
 13  &   5.2825 & 5.2272 & 5.2227 & 5.2239 & 5.2218 & {\bf $-9\times 10^{-4}$} \\
 14  &   5.1860 & 5.1401 & 5.1394 & 5.1417 & 5.1394 & {\bf $-0\times 10^{-4}$} \\
 15  &   5.2498 & 5.2038 & 5.2031 & 5.2058 & 5.2033 & {\bf $ 2\times 10^{-4}$} \\
 16  &   5.2718 & 5.2282 & 5.2272 & 5.2301 & 5.2275 & {\bf $ 3\times 10^{-4}$} \\
 17  &   5.2965 & 5.2563 & 5.2548 & 5.2574 & 5.2550 & {\bf $ 2\times 10^{-4}$} \\
 18  &   5.2194 & 5.1948 & 5.1946 & 5.1953 & 5.1946 & {\bf $ 0\times 10^{-4}$} \\
 19  &   5.2831 & 5.2586 & 5.2583 & 5.2590 & 5.2584 & {\bf $ 1\times 10^{-4}$} \\
\hline \\
\label{S2B8}
\end{tabular}
\end{table}

\begin{table}

\caption{Same as table~\ref{S2B8} but for $S=0$ and $B=5$ T.}

\begin{tabular}{ccccccc}
\hline
  L  & 1LL ED & 2LL ED & 3LL ED & e-e VMC & DMC & DMC-ED \\
\hline
  6  &  4.7787 & 4.5007 & 4.4846 & 4.5041 & 4.4852 & {\bf $ 6\times 10^{-4}$} \\
  7  &  4.7943 & 4.5479 & 4.5365 & 4.5512 & 4.5346 & {\bf $ -9\times 10^{-4}$} \\
  8  &  4.4924 & 4.3872 & 4.3833 & 4.3864 & 4.3825 & {\bf $ -8\times 10^{-4}$} \\
  9  &  4.5875 & 4.4824 & 4.4784 & 4.5075 & 4.4787 & {\bf $ 3\times 10^{-4}$} \\
 10  &  4.4911 & 4.4150 & 4.4131 & 4.4175 & 4.4132 & {\bf $ 1\times 10^{-4}$} \\
\hline \\
\label{S0B5}
\end{tabular}
\end{table}

The energies from 3 LL diagonalizations are indistinguishable from the DMC energies
on the scale of the plots, so we present them as tables.
In Table~\ref{S2B8} we show the energies for $N=4$, $S=2$, $B=8$ T and in Table~\ref{S0B5}
for $N=4$, $S=0$, $B=5$~T. We have also included the results obtained by
including only the electron-electron part of the Jastrow factor.
The electron-electron Jastrow recovers most of the missing energy by keeping
electrons apart, in agreement with the VMC calculations of spin polarized electrons in
Ref.~\onlinecite{harju_1999}.
For spin polarized electrons DMC gives
better energies than 3 LL calculations especially for the lower $L$ values. For higher $L$,
electrons occupy larger orbits, making the electron-electron Jastrow less
important. Thus, the effect of LL mixing is less important and exact diagonalization results agree
with QMC results within a few standard deviations ($\approx 1\times 10^{-4}$ H$^*$)
of the QMC results. For spin unpolarized electrons, while the QMC results are not
particularly superior to 3 LLs results, agreement still remains excellent as seen
in Table~\ref{S0B5}. Since both VMC and DMC calculations are performed within the
fixed phase approximation, we conclude that exact LLL wave functions have very accurate
phases.

\subsection{MDD-LDD transition}

\begin{figure}[htb]
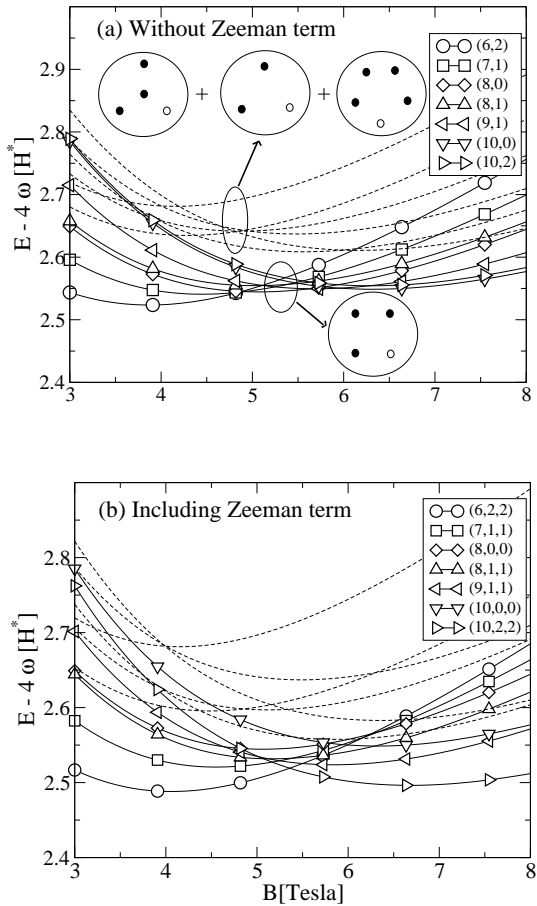

\includegraphics[width=7cm,clip]{fig2a.eps}
%\end{figure}
%\begin{figure}
\vspace*{1cm}

\includegraphics[width=7cm,clip]{fig2b.eps}
\caption{Energy as a function of magnetic field for different many-body states for
$N=4$, (a) without Zeeman term, (b) with Zeeman term. Lines with symbols
represent many-body states which are closely involved in the MDD-LDD transition,
and they are identified by their $(L,S)$ values in (a), and by their $(L,S,S_z)$
values in (b). The locations of the maxima in the
pair densities are indicated by solid dots, and, open dots represent the fixed electrons.}
\label{E_B}
\end{figure}

Having established the accuracy of the form of the wave function,
in this subsection, we study in detail the MDD-LDD transition for $N=4$ and $N=6$.
Fig.~\ref{E_B} shows the energies as a function
of magnetic field for different many-body states $(L,S)$ ($S_z$ is also given in
Fig.~\ref{E_B}(b)) closely involved in MDD-LDD
transition. In Fig.~\ref{E_B}(a), the Zeeman splitting is not taken account, so
the states are $2S+1$-fold degenerate.
It should be noted that, in these calculations only the states with angular momentum
$N(N-1)/2 \leq L \leq N(N+1)/2$ are considered.  For the fully spin polarized case the
MDD-LDD transition involves only the $L=N(N-1)/2$ and $L=N(N+1)/2$ states~\cite{eric_yang_1993,maksym_2000},
but LLL exact diagonalization energies indicate that
other angular momentum states between these values are expected to become ground states
for spin unpolarized or partially polarized electrons~\cite{eric_yang_1993}.
The excitation spectrum in Fig.~\ref{E_B}(a) reveals a very interesting structure, involving several
many-body states in a small region near the MDD-LDD transition which occurs at
$B\approx5.5$ T: While these states (shown by solid lines with symbols)
don't seem to have a regular predictable $(L,S)$ pattern, they correspond to
the magic number sequences of different $S$ states for $N=4$ ~\cite{maksym_2000,kamilla_1995,jain_1995},
Strikingly, the MDD-LDD transition acts as a convergence point for these particular ($L^*,S^*$)
values.  On the other hand, when the Zeeman effect is taken into account
(we consider an effective Land\'e factor $g^*=-0.44$), higher spin states
are favored and the separation between the solid lines with symbols and dashed lines become less
clear.

An interesting property of these states can be
revealed by plotting their pair-density
$\rho({\bf r}_1,{\bf r}_2)$ (probability density of two electrons being at ${\bf r}_1$ and ${\bf r}_2$),
with the reference electron ${\bf r}_2$ fixed at a density maximum.
As schematically shown in Fig.~\ref{E_B}(a), the lower
energy states all have the same structure, with electrons arranged in a square. In fact,
square symmetry corresponds to the minimum energy electronic arrangement of classical
point charges~\cite{maksym_2000}. The higher
energy states (dashed lines) have their electrons arranged in a centered triangle, except
for the two center of mass excited states $(L,S,C_-)=(7,2,1)$ and $(9,0,1)$.
These two states
show 3 and 5 peaks respectively in their pair-densities instead of the more intuitive 4-peak structure
(including the reference electron) one would expect for a $N=4$ dot.
We have also observed similar structures
in other center of mass excitations for higher $L$ values.

\begin{figure}[htb]
\includegraphics[width=8cm,clip]{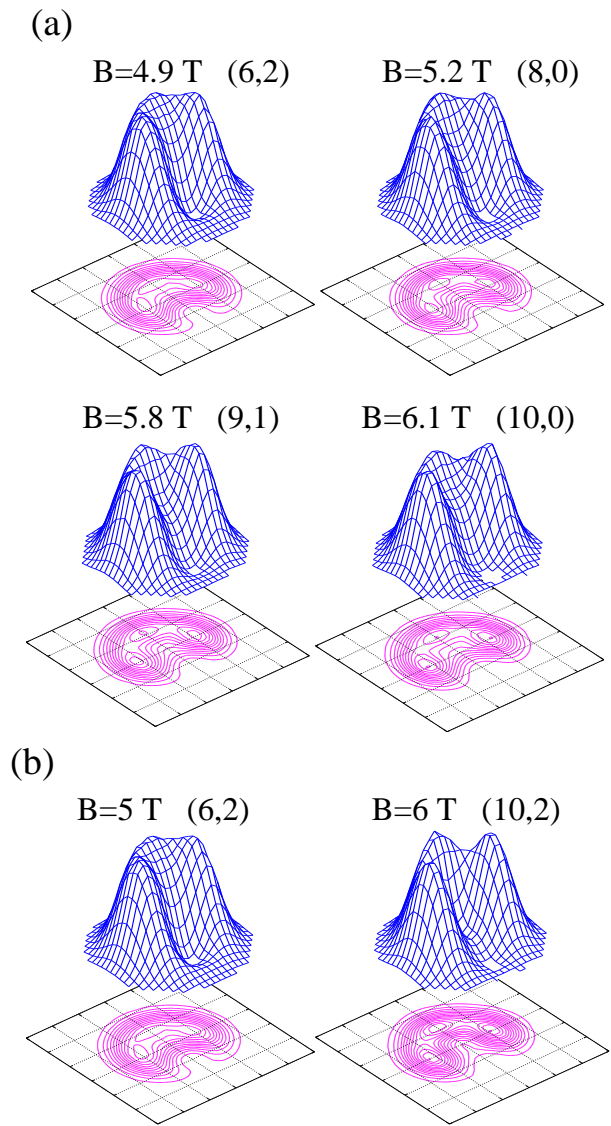}
\caption{(color online) Evolution of VMC ground state pair densities as a function of magnetic field
for $N=4$. (a) Without Zeeman effect (b) Including Zeeman effect.}
\label{rho}
\end{figure}

In Fig.~\ref{rho} we study the ground state pair densities at magnetic fields between $5-6$ T
as the system goes from the MDD state to the LDD states. Without Zeeman splitting, there is a smooth
electronic redistribution involving the states $(6,2)$, $(8,0)$, $(9,1)$, and $(10,0)$.
with increased localization of electrons.
Note that, in this case, the state $(10,2)$ never becomes the ground state in agreement
with the results of Ref.(\onlinecite{wensauer_2004}). When the
Zeeman term is included,
low spin states are suppressed and
the system becomes fully polarized. As a result, the MDD-LDD transition is much more
abrupt, going directly from the state $(6,2)$ to the state $(10,2)$. An abrupt
charge redistribution in post-MDD transitions was observed experimentally also
~\cite{oosterkamp_1999}.
The dominant determinant in the $(10,2)$ state has LLL orbitals with $l=1,2,...N$ and consequently
the density at the center of the dot is small.

\begin{figure}[htb]
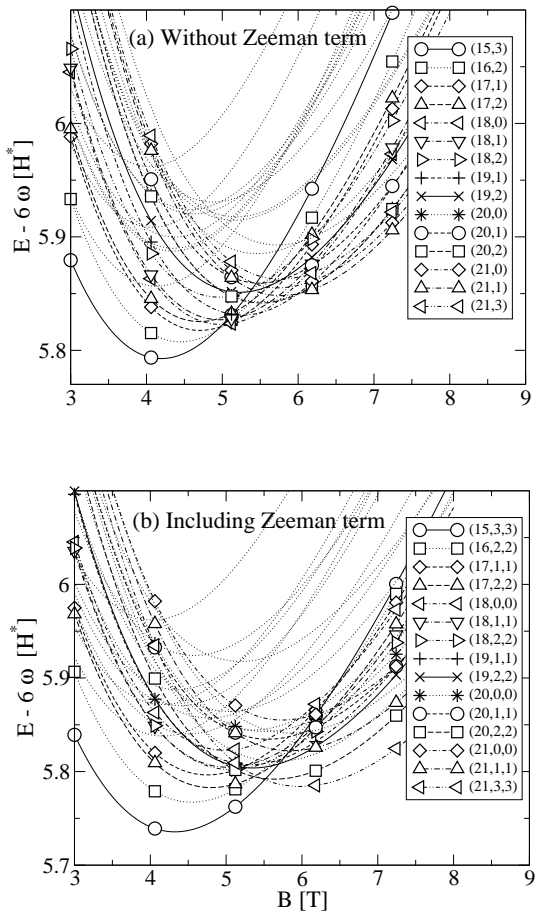

\includegraphics[width=7cm,clip]{fig4a.eps}
\vspace*{1cm}

\includegraphics[width=7cm]{fig4b.eps}
\caption{Energy as a function of magnetic field for different many-body states for
$N=6$, (a) without Zeeman term, (b) with Zeeman term. Lines with symbols
represent many-body states which are closely involved in the MDD-LDD transition,
and they are identified by their $(L,S)$ values in (a), and by their $(L,S,S_z)$
values in (b).}
\label{E_N6}
\end{figure}

\begin{figure}[htb]
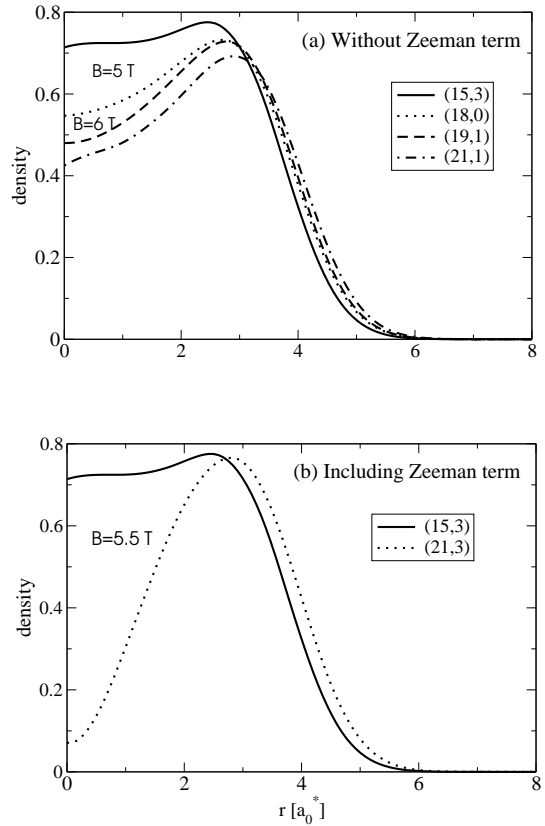

\includegraphics[width=7cm,clip]{fig5a.eps}
\vspace*{1cm}

\includegraphics[width=7cm]{fig5b.eps}
\caption{Evolution of the ground state densities as a functions of magnetic field
is increased for $N=6$, obtained by VMC method. (a) Without Zeeman effect (b) Including
Zeeman effect.}
\label{rho_N6}
\end{figure}

In Fig.~\ref{E_N6}, we plot the energy spectrum for $N=6$. As in the $N=4$ case, we can distinguish
between two sets of eigenstates when the Zeeman term is neglected (Fig.4(a)).
We have only labeled the states that are closely involved in the
MDD-LDD transition, marked by symbols. However, unlike the $N=4$ case, we have
found that, at the confinement strength considered here ($\omega_0=3.32$ meV),
the pair densities of $N=6$ dots near MDD-LDD transition have little structure
(except for the $(21,3)$ state which has a small hexagonal-shaped ripple).
In Fig.~\ref{rho_N6}, we plot the radial densities
of the ground states that participate in the MDD-LDD transition, namely,
$(15,3)$, $(18,0)$, $(19,1)$, and $(21,1)$. These ground state transitions
are in agreement with those of Ref.~\onlinecite{siljamaki_2002} except for the last
transition where they observe a large region where the $(20,2)$ state occurs,
followed by $(21,3)$. In our calculations, as seen in Fig.~\ref{E_N6},
the $(21,3)$ state occurs only if a Zeeman term is included.
As a result of the Zeeman splitting,
the charge redistribution during the MDD-LDD transition
becomes more abrupt, as in the $N=4$ dot.

\section{Conclusions}
To summarize, by combining exact diagonalization and QMC methods, we have obtained
highly accurate many-body wave functions and energies for circular parabolic dots in the
range of magnetic fields corresponding to the MDD-LDD transition.
In this regime, the dot goes through several angular momentum and spin transitions,
and one does not expect Hartree-Fock theory, or, density functional
theory with an approximate functional, to provide an accurate description.
Exact diagonalization is prohibitively expensive because of the
importance of Landau level mixing.
Exact diagonalization wave functions within the LLL approximation are not
sufficiently accurate on their own, but when they are multiplied by an
optimized Jastrow and used as trial wave functions in DMC, they lead to
very accurate energies.
The high accuracy of the fixed phase DMC energies
shows that the phase of exact wave functions is not strongly affected by LL mixing.
The method described is very efficient
since the trial wave functions have determinantal coefficients that are independent of system
parameters $m^*$, $\omega_0$, $B$ and $\epsilon$, and the Jastrow optimization process
is very fast and stable.
We have found that near the MDD-LDD transition, the pair densities for $N=4$
have a well defined square structure, whereas
those for $N=6$ have little structure.
When the Zeeman energy is included, the next ground state beyond the MDD state
is fully polarized and has $L=N(N+1)/2$.  This leads to an abrupt redistribution of
charge, as observed experimentally~\cite{oosterkamp_1999}.  If instead the Zeeman energy is
zero then several unpolarized and partially polarized states become ground states over
a rather small range of magnetic fields and the charge redistribution is smaller and smoother.
This could be verified experimentally.
For instance, if a material for which the $g-$factor can be tuned to a small value is used,
then the transition from unpolarized or partially polarized LDD states to fully polarized ones
could be induced by increasing the effective $g-$factor by using tilted
magnetic fields~\cite{WeisHaugKlitzingPloog93}.

The number of electrons that can be studied with this method is limited by
the number of Slater determinants in the LLL approximation.  This increases
rapidly as the angular momentum increases beyond the MDD value, $L=N(N-1)/2)$.
If the system is fully spin polarized, the number of determinants in the LLL
approximation is much reduced and $N$ can be much larger.
When $L$ is sufficiently large that the number of determinants in even the
LLL becomes too large it is possible to dramatically reduce the number of
determinants by employing
composite fermion wave functions~\cite{JeonChangJain04}, which are very good approximations
to the exact LLL wave functions.
Once again Landau level mixing can be incorporated very efficiently with QMC.
This has been done for a 15-electron dot in Ref.~\onlinecite{GucluGeonUmrigarJain05}.

\section{Acknowledgments:}
We thank W. Geist, J. Jain, M. Korkusinski and F. Pederiva for helpful discussions.
This work was supported by the NSF and FCAR.
\vfill

%%%%%%%%%%%%%%%%%%%%%%%%%%%%%%%%%%%%%%%%%%%%%%%%%%%%%%%%%%%%%%%%%%%%%%%%
%
% BIBLIOGRAPHY:
%
%%%%%%%%%%%%%%%%%%%%%%%%%%%%%%%%%%%%%%%%%%%%%%%%%%%%%%%%%%%%%%%%%%%%%%%%

\end{document}